\documentclass[aps,prl,preprintnumbers,showpacs,nofootinbib]{revtex4}

\usepackage{graphicx}
\usepackage{dcolumn}
\usepackage{bm}
\usepackage{subfig}
\begin{document}
\newcommand{\newc}{\newcommand}
\newc{\ra}{\rightarrow}
\newc{\lra}{\leftrightarrow}
\newc{\beq}{\begin{equation}}
\newc{\eeq}{\end{equation}}
\newc{\barr}{\begin{eqnarray}}
\newc{\earr}{\end{eqnarray}}
\newcommand{\Od}{{\cal O}}
\newcommand{\lsim}   {\mathrel{\mathop{\kern 0pt \rlap
  {\raise.2ex\hbox{$<$}}}
  \lower.9ex\hbox{\kern-.190em $\sim$}}}
\newcommand{\gsim}   {\mathrel{\mathop{\kern 0pt \rlap
  {\raise.2ex\hbox{$>$}}}
  \lower.9ex\hbox{\kern-.190em $\sim$}}}

\title{SEARCHING FOR DARK MATTER-\\
THEORETICAL RATES AND EXCLUSION PLOTS DUE TO THE SPIN
 }

\author{ J. D. Vergados\thanks{Vergados@cc.uoi.gr}}
\affiliation{
{\it Theoretical Physics Division, University of Ioannina,
Ioannina, Gr 451 10, Greece} \\and\\
University of Tuebingen, Tuebingen, Germany.}
\begin{abstract}
The recent WMAP data have confirmed that exotic dark matter
together with the vacuum energy (cosmological constant) dominate
in the flat Universe. The nature of the dark matter constituents
cannot be determined till they are directly detected. Recent
developments in particle physics provide a number of candidates as
constituents of dark matter, called Weakly Interacting Massive
Particles (WIMPs). Since these interact weakly and are of low
energy, they cannot excite the target and can only be detected via
measuring the recoiling nucleus. For all WIMPs, including the most
popular candidate, the lightest supersymmetric particle (LSP), the
relevant cross sections arise out of the following mechanisms: i)
The coherent mode, due to the scalar interaction. ii) The charge
coherent mode, with only proton contribution, as
 in the recent case of secluded dark matter scenario and iii) The spin
contribution arising from the axial current. In this paper we will
focus on the spin contribution, which maybe important, especially
for light targets.

\end{abstract}

\date{\today}
\maketitle
\section{Introduction}
 Combining the
recent WMAP data \cite{WMAP08a} with other experiments one finds  that most of the matter in the Universe is cold dark matter (CDM):
$$\Omega_b=0.0456 \pm 0.0015, \Omega _{CDM}=0.228 \pm 0.013 , \Omega_{\Lambda}= 0.726 \pm 0.015$$

Since the non exotic component cannot exceed $40\%$ of the CDM
~\cite {Benne}, there is room for exotic WIMPs (Weakly
Interacting Massive Particles).
  In fact the DAMA experiment ~\cite {BERNA2} has claimed the observation of one signal in direct
detection of a WIMP, which with better statistics has subsequently
been interpreted as a modulation signal \cite{BERNA1}. These  data,
however, if they are due to the coherent process, are not
consistent with other recent experiments, see e.g. EDELWEISS
\cite{EDELWEISS}, CDMS \cite{CDMS} and XENON \cite{XENON08}. The DAMA results could still be interpreted as due
to the
spin cross section, but with a new interpretation of the extracted
nucleon cross section.

 Supersymmetry naturally provides candidates for the dark matter constituents
(see, e.g, the review \cite{Jung}).
 In the most favored scenario of supersymmetry the
LSP (Lightest Supersymmetric
Particle) can be simply described as a Majorana fermion, a linear
combination of the neutral components of the gauginos and
higgsinos \cite{Jung},\cite{ref2}. In most
calculations the neutralino is assumed to be primarily a gaugino,
usually a bino. Models which predict a substantial fraction of
higgsino lead to a relatively large spin induced cross section due
to the Z-exchange. Such models have been less popular, since they
tend to violate the relic abundance constraint.
These  fairly stringent constrains, however, apply only in the
thermal production mechanism. Furthermore they do not affect the
WIMP density in our vicinity derived from the rotational curves.
 Thus one may assume that large spin cross sections are possible in such models \cite{CHATTO},
\cite{WELLS}, which are non-universal gaugino mass models and give
rise to large higgsino components. Sizable spin cross sections
also arise in the context of other models, which have appeared
recently \cite{JDVSPIN04}, \cite{EOSS04}-\cite{HMNS05} (see also
Ellis {\it et al} \cite{EOSS05} for an  update).\\
 Spin induced cross sections also arize in the case
Kaluza-Klein (K-K) WIMPs in models with Universal Extra
Dimensions (UED) \cite{OikVerMou}. This occurs regardless of
whether the WIMP is a K-K boson or a K-K neutrino. They can also
arise in technicolor theories \cite {KOUVARIS07}. In
the Ultra Minimal Walking Technicolor model
 \cite{GudKouv06,KHLKOUV08,RytSan08} there exist singlet composite
Majorana fermionic states. These, taken as dark matter candidates, can lead to spin induced cross sections.

 Knowledge of the spin induced nucleon cross section is very
 important since, for some special targets, it may lead to transitions to excited nuclear  states,
 which provide the attractive signature of detecting the
 de-excitation $\gamma$ rays in or without coincidence with the
 recoiling nucleus \cite{eji93},\cite{VQS04}. Furthermore it may dominate in  light
 systems like  $^3$He and $^{19}$F, which offer some attractive advantages \cite{SANTOS04},\cite{PICASSO09}.

In light of the above it is clear that the spin mechanism needs be considered. In this article we will discuss the  theoretical ingredients needed to obtain the WIMP-nuclear  spin induced cross sections. Then we will give expressions for and calculate the event rates, both modulated and unmodulated, in terms of the elementary proton ($\sigma_p$) or neutron ($\sigma_n$) cross sections. After that we will provide exlusion plots in the $(\sigma_p,\sigma_n)$ plane in terms of parameters relevant to the experiments, for various targets of experimental interest. Our results will be presented in a way that will make them useful in the analysis of the data of the odd mass targets.
\section{The Essential Theoretical Ingredients  of Direct Detection.}
 Even though there exists firm indirect evidence for a halo of dark matter
 in galaxies from the
 observed rotational curves, it is essential to directly detect
 such matter. Such a direct detection, among other things, may also
 unravel the nature of the constituent of cold dark matter, namely the Weakly Interacting Massive Particle (WIMP).
 The
 possibility of such detection, however, depends on the nature of its
 constituents.
 Our main conclusions apply to all heavy WIMPs.
  Since the WIMP is expected to be very massive, $m_{\chi} \geq 30 GeV$, and
extremely non relativistic with average kinetic energy $T \approx
50KeV (m_{\chi}/ 100 GeV)$, it can be directly detected
 mainly via the recoiling of a nucleus
(A,Z) in elastic scattering. The event rate for such a process can
be computed from the following ingredients:
\begin{enumerate}
\item An effective Lagrangian at the elementary particle (quark)
level obtained in the framework of supersymmetry as described ,
e.g., in Refs~\cite{ref2,JDV96}.  An analogous procedure can be
found in the case of K-K WIMPs in  Universal Extra Dimension
(UED) models \cite{OikVerMou} and technicolor  theories \cite
{KOUVARIS07}.
\item A well defined procedure
for transforming the amplitude obtained using the above mentioned
effective Lagrangian from the quark to the nucleon level
where
 This step is not trivial, since the
obtained results depend crucially on the content of the nucleon in
quarks other than u and d. This is particularly true for the
scalar couplings, which are proportional to the quark
masses~\cite{Dree,Dree,Dree00,Chen} as well as the isoscalar axial
coupling.
 \item Nuclear matrix elements. \\
 These must be obtained with as reliable as possible
many body nuclear wave functions. Fortunately in the most studied
case of the scalar coupling the situation is quite simple, since
then one needs only the nuclear form factor. Some progress has
also been made in obtaining reliable static spin matrix elements
and spin response functions \cite{DIVA00,Ress,SUHONEN03,KVprd}
\item A velocity distribution for WIMPs\\
In this article we will follow the standard practice and assume a M-B distribution, but other perhaps
more realistic velocity distributions have also recently been
considered \cite{VerHanH,JDV09}
\end{enumerate}
Since the obtained rates are very low, one would like to be able
to exploit the modulation of the event rates due to the earth's
revolution around the sun \cite{DFS86,FFG88,Verg98,Verg01}

 \section{THE WIMP NUCLEUS CROSS SECTIONS}
\begin{figure}
\begin{center}
\includegraphics[height=.20\textheight]{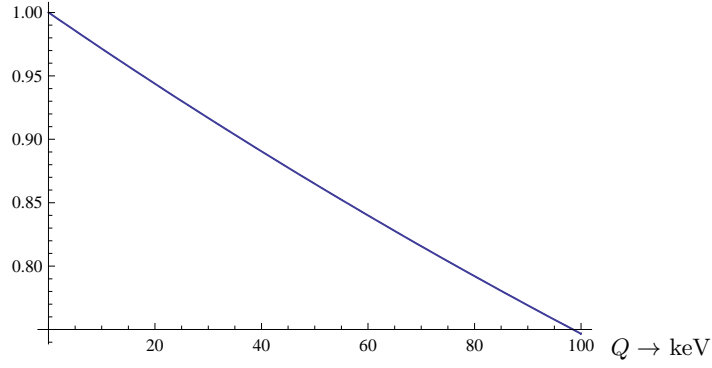}
\hspace{-0.0cm} $Q\rightarrow$ keV
\caption{ The spin response functions $F_{11}(Q)$, $F_{01}(Q)$ and $F_{00}(Q)$ in the case of the target $^{19}$F as a function of the energy transfer. In the region of interest for dark matter searches they are not distinguishable.
 \label{Fig:spinFF19} }
\end{center}
\end{figure}

The standard (non directional)  differential  rate can be written as
\begin{equation}
dR = \frac{\rho (0)}{m_{\chi^0}} \frac{m}{A m_N}
 d\sigma (u,\upsilon) | {\boldmath \upsilon}|,
\label{Eq:difrate}
\end{equation}
where
 m is the detector mass, $\rho (0) = 0.3 GeV/cm^3$ is the WIMP density in our vicinity, $\upsilon$ its velocity
and
 $m_{\chi^0}$ its mass
%
and $d\sigma(u,\upsilon )$ is given by
\beq
d\sigma (u,\upsilon)== \frac{du}{2 (\mu _r b\upsilon )^2}
 \left [\bar{\Sigma} _{S}F^2(u)
                       +\bar{\Sigma} _{spin} F_{11}(u) \right ],
\label{2.9}
\end{equation}
where $ u$ is the energy transfer $Q$ in dimensionless units given by
\begin{equation}
 u=\frac{Q}{Q_0}~~,~~Q_{0}=[m_pAb]^{-2}=40A^{-4/3}~MeV,
\label{defineu}
\end{equation}
 with  $b$ is the nuclear (harmonic oscillator) size parameter. $F(u)$ is the
nuclear form factor and $F_{11}(u)$ is the spin response function
associated with the isovector channel. The scalar cross section is
given by:
\begin{equation}
\bar{\Sigma} _S
\approx  \sigma^{S}_{N,\chi^0} \left (\frac{\mu_r}{\mu_r (p)} \right )^2 A^2.
\label{2.10}
\end{equation}
$\sigma^S_{N,\chi^0}$ is the WIMP-nucleon scalar
cross section. Note that, since the heavy quarks dominate, the isovector contribution is
negligible, i.e. the proton and nucleon cross sections are the same. The spin Cross section is given by:
\begin{equation}
\bar{\Sigma} _{spin}  =  (\frac{\mu_r}{\mu_r(p)})^2
                           \sigma^{spin}_{p,\chi^0}~\zeta_{spin},
\zeta_{spin}= \frac{1}{3(1+\frac{f^0_A}{f^1_A})^2}S(u),
\label{2.10a}
\end{equation}
\begin{equation}
S(u)=\left [\left (\frac{f^0_A}{f^1_A} \Omega_0(0)\right )^2\frac{F_{00}(u)}{F_{11}(u)}
  +  2\frac{f^0_A}{ f^1_A} \Omega_0(0) \Omega_1(0)\frac{F_{01}(u)}{F_{11}(u)}+  \Omega_1(0))^2  \, \right ].
\label{s(u)}
 \end{equation}
The spin response functions $ F_{ij}$, properly normalized to unity at momentum transfer zero, in the energy transfers of interest are almost the same for
all isospin channels $i,j;i,j=0,1$ . As a matter of fact in the case of $^{19}$F we get:
\beq
F_{11}(u)=e^{-u} \left(0.0119 u^4-0.1450 u^3+0.6620
   u^2-\frac{4 u}{3}+1\right),
\eeq
\beq
F_{01}=e^{-u} \left(0.0124 u^4-0.1487 u^3+0.6677 u^2-\frac{4 u}{3}+1\right),
\eeq
\beq
F_{00}=e^{-u} \left(0.0131 u^4-0.1525 u^3+0.6733 u^2-\frac{4 u}{3}+1\right),
\eeq
Thus they are indistinguishable (see Fig. \ref{Fig:spinFF19}) for the energy transfers of interest. Hence

\begin{equation}
S(u)\approx S(0)=\left (\frac{f^0_A}{f^1_A}\Omega_0(0)+\Omega_1(0) \right )^2.
\label{s(0)}
 \end{equation}
 The nuclear matrix elements $\Omega_1(0)$ ($\Omega_0(0)$) associated
 with the isovector (isoscalar) components are normalized so that, in the case
 of the proton, they yield $\zeta_{spin}=1$  at $u=0$.\\
 The couplings $f^1_A$ ($f^0_A$) are obtained by multiplying the corresponding
elementary amplitudes obtained at the quark level by suitable
renormalization factors $g^0_A$ and $g^1_A$ given in terms of the
quantities $\Delta q$ given by Ellis \cite{JELLIS} \beq
g_A^0=\Delta u+\Delta d+\Delta
s=0.77-0.49-0.15=0.13~,~g_A^1=\Delta u-\Delta d=1.26 \eeq Thus,
barring very unusual circumstances at the quark level, the
isovector component is expected to be dominant.
It is for this reason that we started our discussion in the isospin basis.\\
Heavy nuclei, however, are theoretically described in terms of protons and
neutrons and the experiments are also analyzed this way.
So will present our results in this basis.
 The proton and  neutron cross section are given by:
 \beq
\sigma^{spin}_{p,\chi^0}=3 \sigma_0 |{f^0_A}+{f^1_A}|^2=3 \sigma_0|a_p|^2~,~\sigma^{spin}_{n,\chi^0}=3 \sigma_0 |{f^0_A}-{f^1_A}|^2=3 \sigma_0|a_n|^2
\label{eq:eg 52}
\eeq
with $a_p$ and $a_n$ are the proton and neutron spin amplitudes, which, of course, depend on the model.
In the case of the LSP \cite{JDV96}
$$\sigma_0 = \frac{1}{2\pi} (G_F m_p)^2 = 0.77 \times 10^{-38}
cm^2 = 0.77 \times 10^{-2} pb.$$
 \\In extracting limits on the nucleon cross sections from the data we will find it convenient to write:
 \begin{equation}
\bar{\Sigma} _{spin}  =  (\frac{\mu_r}{\mu_r(p)})^2 \sigma^{spin}_{nuc}~,~\sigma^{spin}_{nuc}=
\frac{1}{3}|\Omega_p \sqrt{ \sigma_p}+{\Omega_n} \sqrt{\sigma_n}
 e^{i \delta}|^2=\frac{1}{3}||\Omega_p| \sqrt{ \sigma_p}+|{\Omega_n} |\sqrt{\sigma_n}
 e^{i \left(\delta+\delta_A\right )}|^2,
\label{2.10ab}
\end{equation}
 where $\Omega_p(0)$ and $\Omega_n(0)$ are the proton and neutron components of the static spin nuclear matrix elements, $\delta_A$ is the relative phase between them (zero or $\pi$) and
  $\delta$ the relative
phase between the  amplitudes $a_p$ and $a_n$.\\
The nuclear spin ME are defined as follows:
\beq
\Omega_p(0)=\sqrt{\frac{J+1}{J}}\prec J~J| \sigma_z(p)|J~J\succ ~~,~~
\Omega_n(0)=\sqrt{\frac{J+1}{J}}\prec J~J| \sigma_z(n)|J~J\succ
\label{Omegapn}
\eeq
where $J$ is the total angular momentum of the nucleus and $\sigma_z=2 S_z$. The spin operator is defined by
$S_z(p)=\sum_{i=1}^{Z} S_z(i)$, i.e. a sum over all protons in the nucleus,  and
$S_z(n)=\sum_{i=1}^{N}S_z(i)$, i.e. a sum over all neutrons. Furthermore
\beq
\Omega_0(0)=\Omega_p(0)+\Omega_n(0)~~,~~
\Omega_1(0)=\Omega_p(0)-\Omega_n(0).
\label{Omegaiso}
\eeq
The spin ME can be  obtained in the context of a given nuclear model. Some such matrix elements of interest to the planned
experiments are given in table \ref{table.spin}.
 The shown results
are obtained from DIVARI \cite{DIVA00}, Ressel {\it et al} (*) \cite{Ress},
 the Finish group (**) \cite {SUHONEN03} and the Ioannina team (+) \cite{ref1}, \cite{KVprd}.

\begin{table}[t]
\caption{
 The static spin matrix elements for various nuclei. For $^3$He see Moulin, Mayet and Santos
\cite{Santos}. For the other
light nuclei the calculations are from DIVARI \cite{DIVA00}.
 For  $^{73}$Ge and $^{127}$I the results presented  are from Ressel {\it et al}
\cite{Ress} (*) and the Finish group {\it et al} \cite {SUHONEN03}
 (**).
 For $^{207}$Pb they were obtained by the Ioannina team (+).
\cite{ref1}, \cite{KVprd}.
 \label{table.spin} }
\begin{center}
\begin{tabular}{lrrrrrrrr}
\hline\hline
 &   &  &  &  &   &  & &\\
 &$^3$ He& $^{19}$F & $^{29}$Si & $^{23}$Na  & $^{73}$Ge & $^{127}$I$^*$ & $ ^{127}$I$^{**}$ & $^{207}$Pb$^+$\\
\hline
    &   &  &  &  &    \\
$\Omega_{0}(0)$ &1.244     & 1.616   & 0.455  & 0.691  &1.075 & 1.815  &1.220  & 0.552\\
$\Omega_{1}(0)$&-1.527     & 1.675  & -0.461  & 0.588 &-1.003 & 1.105  &1.230  & -0.480\\
$\Omega_{p}(0)$ &-0.141    & 1.646  & -0.003  & 0.640  &0.036 &1.460   &1.225  & 0.036\\
$\Omega_{n}(0)$ &1.386     & -0.030   & 0.459  & 0.051  &1.040 & 0.355  &-0.005 & 0.516\\
$\mu_{th} $& &2.91   &-0.50  & 2.22  &    & &\\
$\mu_{exp}$& &2.62   &-0.56  & 2.22  &    & &\\
$\frac{\mu_{th}(spin)}{ \mu_{exp}}$& &0.91   &0.99  & 0.57  &    &  &\\
\hline
\hline
\end{tabular}
\end{center}
\end{table}
Before concluding this section we should emphasize that from the spin matrix elements of Table \ref{table.spin}
 those associated with $^{19}$F are the most reliable for the following reasons \cite{DIVA00}:
\begin{itemize}
\item The light s-d nuclei are very well described within the interacting shell model.
\item The magnetic moment of the ground state is dominated by the spin (the orbital part is negligible).
\item The calculated magnetic moment is quite large and in good agreement with experiment.
\end{itemize}
To summarize: The proton and neutron spin cross sections can be
obtained in a given particle model for the WIMP's.
 As we have seen there is a plethora of such models to motivate the experiments. Some of them may yield  as high as a few tens of events per kg of target per year  \cite{JDVSPIN04}. But most of them depend on imput parameters that are not well detemined. So none of them seems to be
 universally accepted. Thus in the present work, rather than following the standard procedure of providing constrained parameter spaces, we will treat
 the proton and neutron cross sections as parameters to be extracted from the data. This can be done, once the nuclear spin matrix
 elements are known, for various values of the phase difference $\delta$. The only particle parameter we will
 retain is the WIMP mass, which is the most important, since it enters not only in the elementary cross sections but the kinematics as well.
\section{Expressions for the rates and some results}
To obtain the total rates one must fold the diffrential rate of Eq. (\ref{Eq:difrate}) with WIMP velocity and
then integrate  the resultin  expression  over the energy transfer from
$Q_{min}$ determined by the detector energy cutoff to $Q_{max}$
determined by the maximum WIMP velocity (escape velocity, put in
by hand in the M-B distribution), i.e.
$\upsilon_{esc}=2.84~\upsilon_0$ with  $\upsilon_0$ the velocity
of the sun around the center of the galaxy($229~Km/s$).

Ignoring the motion of the Earth the total (non directional) rate
is given by
\begin{equation}
R =  \bar{R}\, t(a,Q_{min})~~,~~
 \bar{R}=\frac{\rho (0)}{m_{\chi^0}} \frac{m}{Am_p}~
              (\frac{\mu_r}{\mu_r(p)})^2~ \sqrt{\langle
v^2 \rangle } \left [\sigma_{p,\chi^0}^{S}~A^2+
 \sigma^{spin} _{nuc} \right ].
 \label{3.55f}
\end{equation}
 The WIMP parameters have been absorbed in $\bar{R}$. The
 parameter $t$ takes care of the nuclear form factor and the
 folding with WIMP velocity distribution \cite{Verg00,Verg01,JDVSPIN04}
 (for its values see table \ref{table.murt}). It depends on
$Q_{min}$, i.e.  the  energy transfer cutoff imposed by the
detector and $a=[\mu_r b \upsilon _0 \sqrt 2 ]^{-1}$.\\
  In the present work  we find it convenient to re-write it as:
\begin{equation}
R= \tilde{K}(\sigma_1) \left[
c_{coh}(A,\mu_r(A),m_{\chi^0})\frac{\sigma_{p,\chi^0}^{S}}{\sigma_1}
+c_{spin}(A,\mu_r(A),m_{\chi^0})\frac{\sigma_{nuc}^{spin}}{\sigma_1} \right]
\label{snew}
\end{equation}
 For the spin cross section it is convenient to take  $\sigma _1=10^{-5}pb$. Thus
\beq
\tilde{K}(\sigma_1)=\frac{\rho (0)}{100\mbox{ GeV}} \frac{m}{m_p}~
              \sqrt{\langle v^2 \rangle }~\sigma_1 \simeq 1.60~10^{-2}~ y^{-1}\frac{\rho(0)}{0.3GeVcm^{-3}}
\frac{m}{1Kg}\frac{ \sqrt{\langle
v^2 \rangle }}{280kms^{-1}}.
\label{Kconst}
\eeq
For the coherent mode it may be more convenient to pick $\sigma _1=10^{-7}pb$, which is close to the present
 experimental limit.
Furthermore
\begin{equation}
c_{coh}(A, \mu_r(A),m_{\chi^0}))=\frac{100\mbox{ GeV}}{m_{\chi^0}}\left[ \frac{\mu_r(A)}{\mu_r(p)} \right]^2 A~t_{coh}(A)~,~
c_{spin}(A, \mu_r(A),m_{\chi^0}))=\frac{100GeV}{m_{\chi^0}}\left[ \frac{\mu_r(A)}{\mu_r(p)} \right]^2 \frac{t_{spin}(A)}{A}.
\label{ctm}
\end{equation}
The parameters $c_{coh}(A,\mu_r(A),m_{\chi^0})$, $c_{spin}(A,\mu_r(A),m_{\chi^0})$, which give the relative merit
 for the coherent and the spin contributions in the case of a nuclear
target compared to those of the proton,  are tabulated in table
\ref{table.murt}
 for
energy cutoff $Q_{min}=0,~10$ keV.
Thus via  Eq. (\ref{snew}) we can  extract the nucleon cross section from
 the data.\\
\begin{table}[t]
\caption{
The factors $c19= c_{coh}(19,\mu_r(19),m_{\chi^0})$,  $s19= c_{spin}(19,\mu_r(19),m_{\chi^0})$,
$c19= c_{coh}(73,\mu_r(73),m_{\chi})$,  $s73= c_{spin}(73,\mu_r(73),m_{\chi})$
 and
$c127=c_{coh}(127,\mu_r(127),m_{\chi^0})$,  $s127= c_{spin}(127,\mu_r(127),m_{\chi^0})$
for two values of $Q_{min}$. Also given are the factors $s3= c_{spin}(3,\mu_r(3),m_{\chi})$ for  $Q_{min}=0$.
\label{table.murt}
}
\begin{center}
\begin{tabular}{|r|r|rrrrrrrr|}
\hline
$Q_{min}$& &\multicolumn{8}{c|}{$m_{\chi}$ (GeV)}\\
\hline
& &   &  &  &  &   & & &\\
keV& & 20 & 30 & 40  & 50 & 60 & 80&100&200\\
\hline
0&t(3,s)&1.166&1.166&1.166&1.166&1.166&1.166&1.166&1.166\\
\hline
0&c3&131& 92.6& 71.6& 58.3& 49.2& 37.5& 30.3&  15.4\\
\hline
0&s3&14.6& 10.3& 7.95& 6.48& 5.47& 4.16& 3.36&  1.71\\
\hline
0&t(19,c)&1.153&1.145&1.138&1.134&1.130&1.124&1.121&1.112\\
0&t(19,s)&1.132&1.117&1.105&1.096&1.089&1.079&1.072&1.056\\
\hline
0&c19&11500& 10500& 9420& 8500& 7700& 6450& 5540& 3210\\
0&s19&31.2& 28.3& 25.4& 22.8& 20.6& 17.2& 14.6& 8.40\\
\hline
\hline
0&t(23,c)&1.107&1.099&1.092&1.089&1.085&1.079&1.076&1.068\\
0&t(23,s)&1.075&1.061&1.050&1.041&1.035&1.025&1.018&1.003\\
\hline
0&c23&16100& 15200& 14100& 1300& 11900& 10200&8830& 5280\\
0&s23&29.5& 27.8& 25.6& 23.4& 21.4& 18.2& 13.8& 9.45\\
\hline
0&t(73,c)& 1.119& 1.083& 1.047& 1.014& 0.984& 0.933& 0.893& 0.780\\
0&t(73,s)& 1.135& 1.112& 1.088& 1.064& 1.043& 1.006& 0.976&0.886\\
\hline
0&c73&113000& 131000& 139000& 143000& 142000& 187000&130000 & 935000\\
0&s73&20.8& 23.9& 25.2& 25.5& 25.2& 23.9& 22.2& 15.4\\
\hline
0&t(127,c)&0.984&0.844&0.721&0.621&0.542&0.430&0.358&0.213\\
0&t(127,s)&0.948&0.796&0.671&0.574&0.501&0.401&0.340&0.220\\
\hline
0&c127& 206000& 225000& 223000& 211000& 197000& 169000& 145000&  82400\\
0&s127&12.3& 13.1& 12.8& 12.1& 11.3& 9.7& 8.5& 5.3\\
\hline
10&t(19,c)&0.352&0.511&0.592&0.639&0.667&0.710&0.720&0.773\\
10&t(19,s)&0.340&0.489&0.563&0.606&0.631&0.669&0.676&0.720\\
\hline
10&c19&3500& 4676& 4902& 4789& 4546& 4075& 3557& 2233\\
10&s19& 9.3& 12.4& 12.9& 12.6& 11.9& 10.6& 9.3& 5.8\\
\hline
 10&t(73,c)&0& 0.020& 0.119& 0.246& 0.363& 0.539& 0.651& 0.847\\
10&t(73,s)&0& 0.0175& 0.105& 0.213& 0.311& 0.453& 0.539& 0.677\\
\hline
10&c73)&0& 2310& 15300& 32900& 49600& 73400& 86300&  89300\\
10&s73&0& 0.39& 2.5& 5.3& 7.9& 11.6& 13.4& 13.4\\
\hline
10&t(127,c)&0.000&0.156&0.205&0.222&0.216&0.191&0.175&0.109\\
10&t(127,s)&0.000&0.135&0.177&0.192&0.190&0.174&0.165&0.121\\
\hline
10&c127& 0& 41500& 63200& 75500& 78500& 74900& 71000& 42200\\
10&s127& 0.& 2.2& 3.4& 4.0& 4.3& 4.2& 4.1& 2.9\\
\hline
\end{tabular}
\end{center}
\end{table}

\begin{figure}
\begin{center}
\includegraphics[height=.20\textheight]{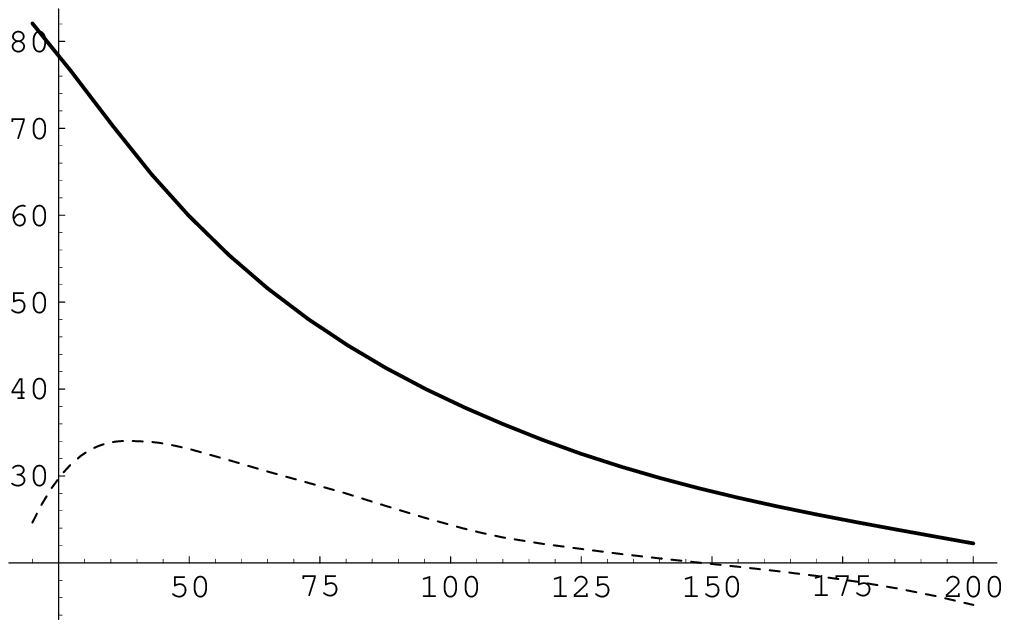}
\includegraphics[height=.20\textheight]{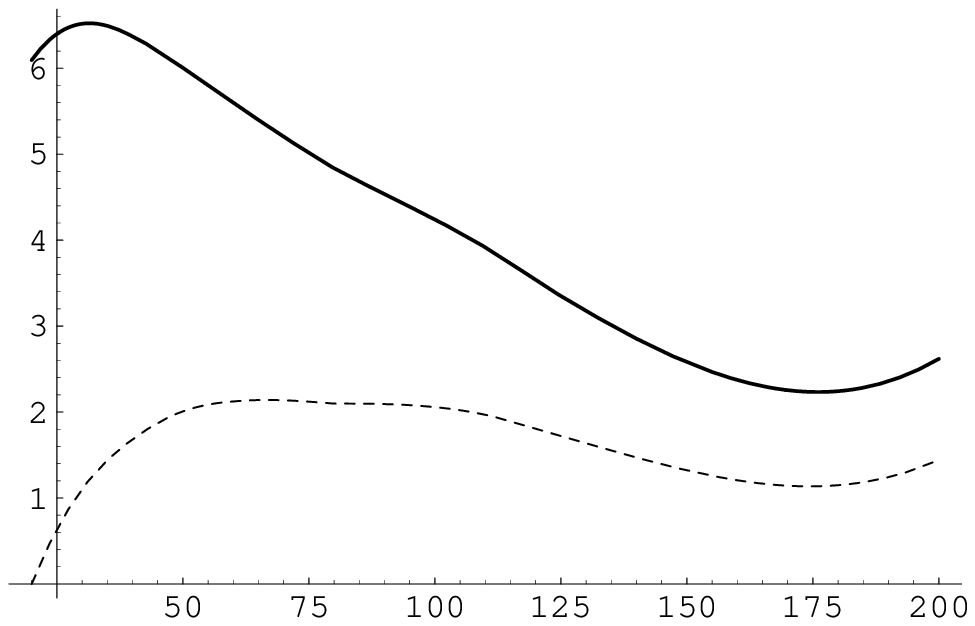}
\caption{ On the left the quantity $c_{spin}(A,\mu_r(A),m_{\chi^0})~\zeta_{spin}$ for the A=19 system is
 shown for two cut off values $Q_{min}=0$,
continuous curve, and $Q_{min}=10$ keV, dotted curve. On the right
the same quantity is shown for the A=127 system. The advantages of
the lighter target, especially for light WIMP, are obvious.
 \label{nume} }
\end{center}
\end{figure}

The  quantity $c_{spin}(A,\mu_r(A),m_{\chi^0})~\zeta_{spin}$, when the
isoscalar contribution is neglected and employing $\Omega ^2_1=
1.22~ (2.81)$ for $^{127}I$ $(^{19}F)$, is shown in Fig
\ref{nume}. In the case of the spin induced cross section, the
light nucleus $^{19}$F has certainly an advantage over the heavier
nucleus $^{127}$I (see Fig. \ref{nume}). For the coherent process,
however, the light nucleus is disfavored.
 (see Table \ref{table.murt}).

The experimental sensitivity ratios (ESR), i.e. the extracted from experiment nucleon cross section ratios satisfy:
\begin{equation}
ESR=\frac{\sigma^{spin}_{k,\chi^0}}{\sigma^{S}_{p,\chi^0}} =
 \left[\frac{c_{coh}(A,\mu_r(A),m_{\chi^0})}{c_{spin}(A,\mu_r(A),m_{\chi^0})}\right]
\frac{3 }{\Omega^2_k}~,k=p,n,iv, \mbox{for proton, neutrom, isovector respectively}
\label{ratior2}
\end{equation}
The quantity $ESR$ for a WIMP mass of 50 GeV is shown in table \ref{table.ESR}.
\begin{table}[t]
\caption{The experimental sensitivity ratios (ESR) for various targets assuming a WIMP mass of 50 GeV. $p,n$
and $iv$ correspond to the elementary proton, neutron and isovector dominance respectively.
\label{table.ESR} }
\begin{center}
\begin{tabular}{||l|r|r|r|r|r||}
\hline\hline
 &   &  &  &  &     \\
 &$^3$ He& $^{19}$F & $^{23}$Na  & $^{73}$Ge & $^{127}$I \\
\hline
    &   &  &  &  &    \\
$p$ &1.4$\times 10^3$     & 4.1$\times 10^2$      & 2.7$\times 10^3$     & 1.3$\times 10^7$     &2.5$\times 10^4$ \\
\hline
$n$ &1.4$\times 10$     & 1.2$\times 10^6$      & 4.3$\times 10^5$     & 1.6$\times 10^4$     &4.2$\times 10^5$ \\
\hline
$iv$ &1.1$\times 10$     & 4.0$\times 10^2$      & 3.2$\times 10^3$     & 1.7$\times 10^4$     &4.3$\times 10^4$ \\
\hline
\hline
\end{tabular}
\end{center}
\end{table}
It is clear from this table why  the limits on the spin  cross section extracted from all
targets is much bigger compared to that extracted for the coherent mode. We should emphasize that the elementary
cross sections do not depend on the target.
It  is only the values extracted from experiment that do so, giving a measure of the sensitivity of the various
experiments. The elementary cross sections only depend on the particle model and the structure of the nucleon.
 Thus, e.g., in the case of K-K WIMPs the coherent cross section dominates, if the WIMP is a K-K gauge boson,
  but the spin cross section is bigger, when the WIMP is a K-K neutrino \cite{OikVerMou}.

\begin{figure}
\begin{center}
\includegraphics[height=.20\textheight]{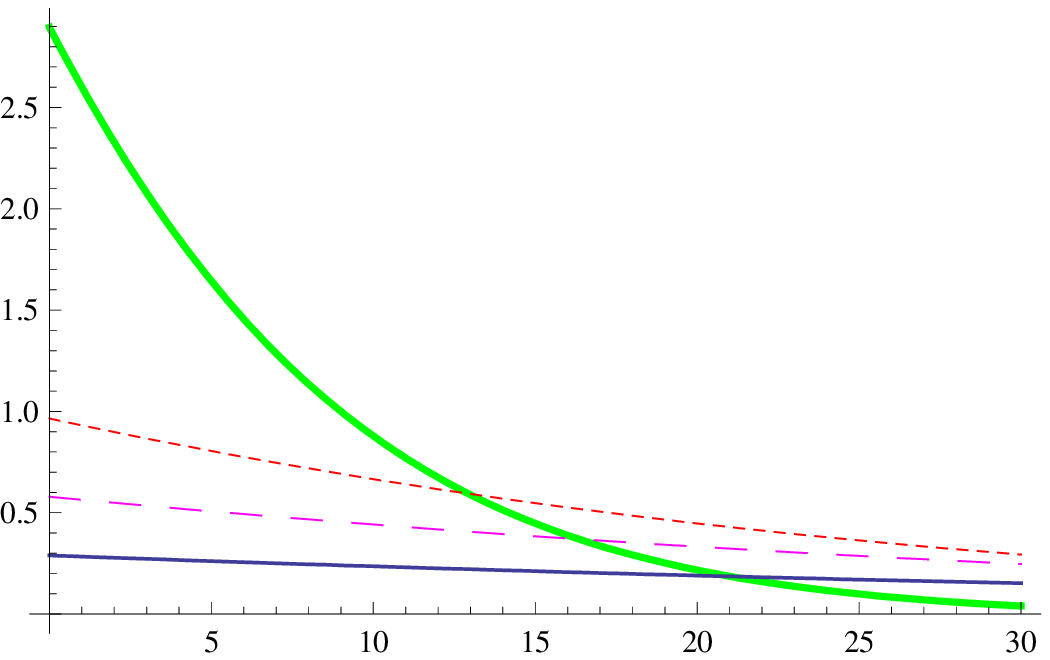}
\includegraphics[height=.20\textheight]{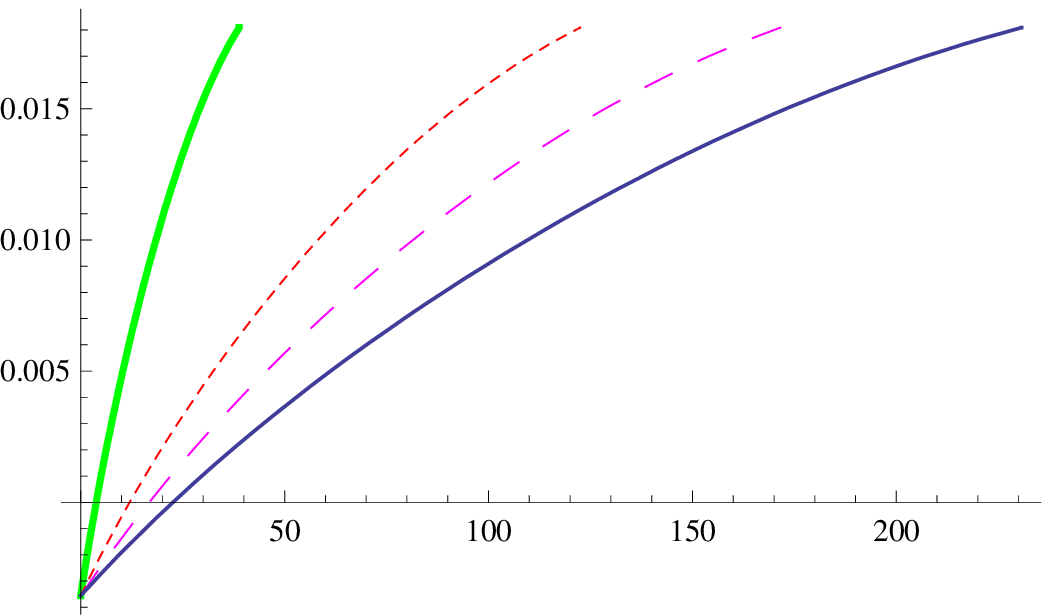}\\
\hspace{-0.0cm} $Q\rightarrow$ keV
\caption{ On the left the quantity $\frac{dR_0(Q,A,m_{\chi})}{dQ} $ and on the right the quantity
 $H(Q,A,m_{\chi}$ involving  the spin induced process for  the A=19 system as a function of the energy
 transfer $Q $ in keV. The thick solid, the dotted , the dashed and the thin solid lines correspond to
  WIMP masses 10,30, 50 and 100 GeV respectively.  On the left panel the range of $Q$ is restricted to make
  the curve for low WIMP mass more visible. For masses heavier than 30 GeV  the differential event rate has
  essentially a constant slope. So it is adequate to restrict ourselves to low $Q$. The full range of Q
  can be inferred from the right panel.
 \label{fig:dRdQH} }
\end{center}
\end{figure}

If the effects of the motion of the Earth around the sun are included, the total
 non directional rate is given by
\begin{equation}
R= \tilde{K}(\sigma_1) \left[c_{coh}(A,\mu_r(A),m_{\chi^0}) \frac{\sigma_{p,\chi^0}^{S}}{\sigma_1}(1 +
h(a,Q_{min})cos{\alpha})\right] \mbox{ (coherent)},
\label{3.55j}
\end{equation}
\begin{equation}
R= \tilde{K}(\sigma_1) \left[c_{spin}(A,\mu_r(A),m_{\chi^0}) \frac{\sigma^{spin}_{nuc}}{\sigma_1}(1 +
 h_{spin}(a,Q_{min})cos{\alpha})\right] \mbox{ (spin)},
\label{3.55jspin}
\end{equation}
where $h$ ($h_{spin}$)  are the modulation amplitudes and
 $\alpha$ is the phase of the Earth, which is
zero around June 2nd. We are  going to only briefly discuss the
modulation  amplitudes here since they depend only on the WIMP
mass and are independent of the other particle parameters. In the
case of the two very light targets, however, they are pretty
independent of the WIMP mass. In fact for the light systems :
\beq~h=h_{spin}=0.0232 \mbox{ (for A=3) and } h=0.0229, h_{spin}=0.0227 \mbox{ (for A=19).  }
\eeq
Actually for the A=19 system there is about (10$\%$) reduction as the WIMP mass increases.
\\In the case of the target $^{3}He$ the quantity $t$ is also  essentially independent of the WIMP mass, since the
 WIMP is expected to be much heavier than the nuclear mass.
 From table \ref{table.murt} we see that the
 coherent rate is quite small for this light system, but the spin induced rate is only a factor of two smaller than
 that for $^{19}$F.  As we have already mentioned this nucleus, has definite experimental advantages \cite{Santos}.
\\In many instances the
experiments are interested in the differential event rate. This is
a function of two variables, the WIMP mass and the energy transfer
$Q$. For the light systems, however, the dependence on the WIMP
mass is rather weak, especially for heavy WIMPs. Thus the
presentation of the results is relatively simple and we are going
to present them here. One finds:
\begin{equation}
\frac{dR}{dQ}= \tilde{K}(\sigma_1) \left[\frac{dR_0(Q,A,m_{\chi})}{dQ} \left (1+H\left (Q,A,m_{\chi}\right )
cos{\alpha} \right )\frac{\sigma^{spin}_{nuc}}{\sigma_1} \right],
\label{3.55j1spin}
\end{equation}
with an analogous expression for the coherent mode. The time average quantity  ${dR_0(Q,A,m_{\chi})}/{dQ} $
 and the relative modulation amplitude $H(Q,A,m_{\chi})$ are shown in Fig. \ref{fig:dRdQH}. The  differential
 cross section is normalized so that the area under the corresponding curve  gives the value $c_{spin}(A,\mu_r(A),m_{\chi^0})
 $ of Eq. (\ref{3.55jspin}). Note that the quantity H, being the  ratio of two amplitudes, the  amplitude for
 modulation divided by the  time independent amplitude, is independent of the nuclear model. So it is the
 same for the spin and coherent mode. Note also that, at relatively low energy transfers, $H$ becomes negative,
 i.e. minimum in June and maximum in December. The negative value, however, for the light targets is small for
 all WIMP masses. For this reason for a light target the integrated modulation amplitude $h$ is always positive
 (maximum in June, minimum in December, as expected). The modulation curves $H $ keep increasing as the energy
 transfer increases, mainly because the time independent amplitude, coming in the denominator, decreases.
 Thus in spite of this increase of $H$, $h$ remains constant.
\section{Results for the Spin Contribution }
\label{results}
\subsection{One amplitude is dominant}
\begin{figure}
\begin{center}
\includegraphics[height=.15\textheight]{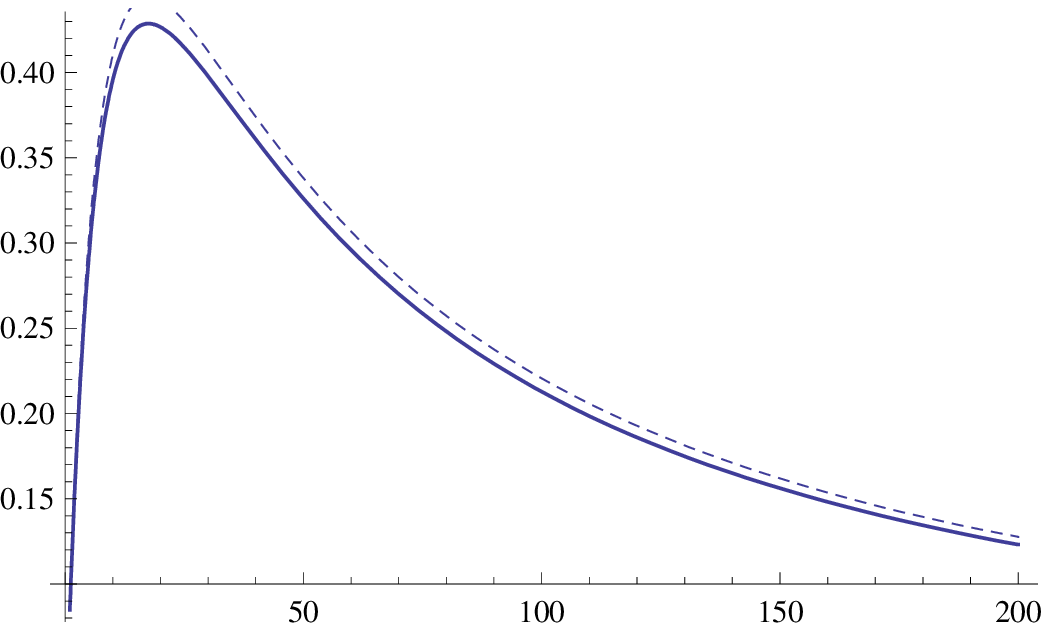}
\hspace{-0.0cm} $m_{\chi}\rightarrow$ GeV
\includegraphics[height=.15\textheight]{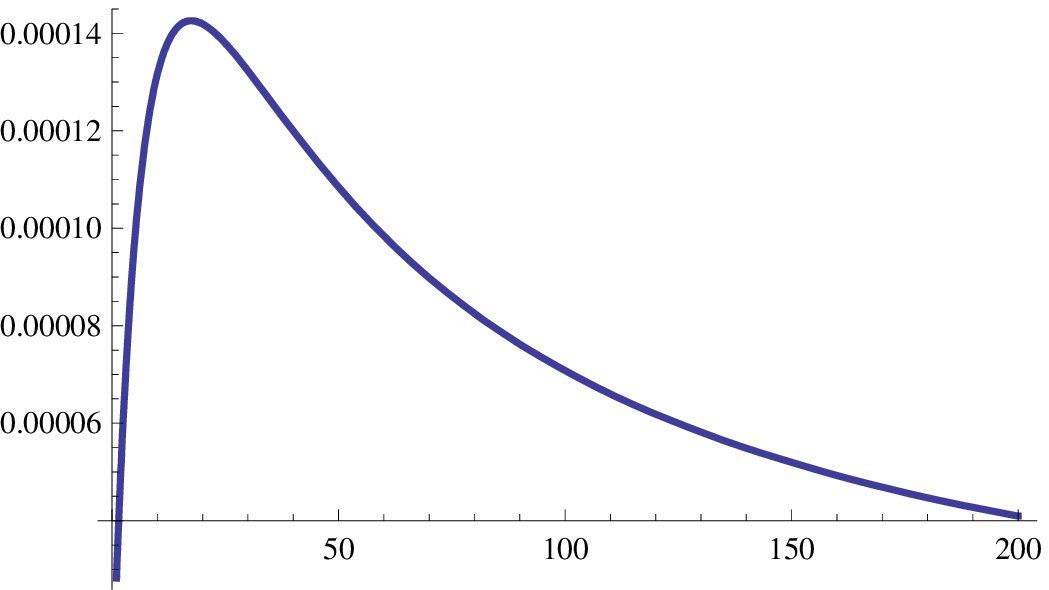}
\hspace{-0.0cm} $m_{\chi}\rightarrow$ GeV
\caption{ The event rate (kg-y) for the target  $^{19}$F assuming a nucleon cross section of $10^{-5}$ pb
as a function of the WIMP mass in GeV. In the left panel the continuous curve takes into account only the
proton component. The dotted curve results when  the proton and neutron cross sections are the same, but
the corresponding amplitudes are opposite (the isoscalar amplitude is assumed to vanish). The difference is small.
In the right panel we consider the case that the elementary proton cross section vanishes. In this case the nuclear
structure suppresses the rate.
 \label{Fig:F19rates}}
\end{center}
\end{figure}

This occurs in cases when the nuclear structure leads to a
dominant spin ME, like $^{19}$F with a dominant proton component.
In this case, barring unusual circumstances  at the quark level
favoring the component not favored by nuclear physics, the
analysis is simple. Thus, e.g., in the case of $^{19}$F
($\Omega_p=1.645$, $\Omega_n=-0.030$ the event rate for an
elementary cross section of $10^{-5}$pb is exhibited as a function
of the WIMP mass in Fig. \ref{Fig:F19rates}. From these plots, for
a given WIMP mass, one may extract limits on the relevant nucleon
cross section from the experimental limits. Using the event rate
of 13.75 Kg-d or 5020 Kg-y of PICASSO \cite{PICASSO09} and the
most favorable WIMP mass of 30 GeV, from Fig. \ref{Fig:F19rates}
we extract a proton spin cross section of 0.1 pb, to be  compared with the
value of 0.16 pb extracted there  \cite{PICASSO09}. From Eq.
(\ref{ratior2}) we extract a coherent cross section of $2.5\times
10^{-4}$pb for this system, which is poor compared to the limits
of CDMS \cite{CDMS} and XENON \cite{XENON08}. The PICASSO people
are fighting with their new detector against the $\alpha$
background, with a flat plateau in the region of their signal, and
their limit will soon substantially improve .
\subsection{Exclusion plots in the  $\sqrt{\sigma_p},\sqrt{\sigma_n}$ plane}
\label{exclplots}
\begin{figure}
\begin{center}
 \subfloat[]
 {
\rotatebox{90}{\hspace{1.0cm}{$ \sqrt{\sigma_n} \rightarrow \sqrt{\sigma_1 \frac{3R}{\tilde{K}(\sigma_1)
|\Omega_n |sA}}$}}
\includegraphics[height=.20\textheight]{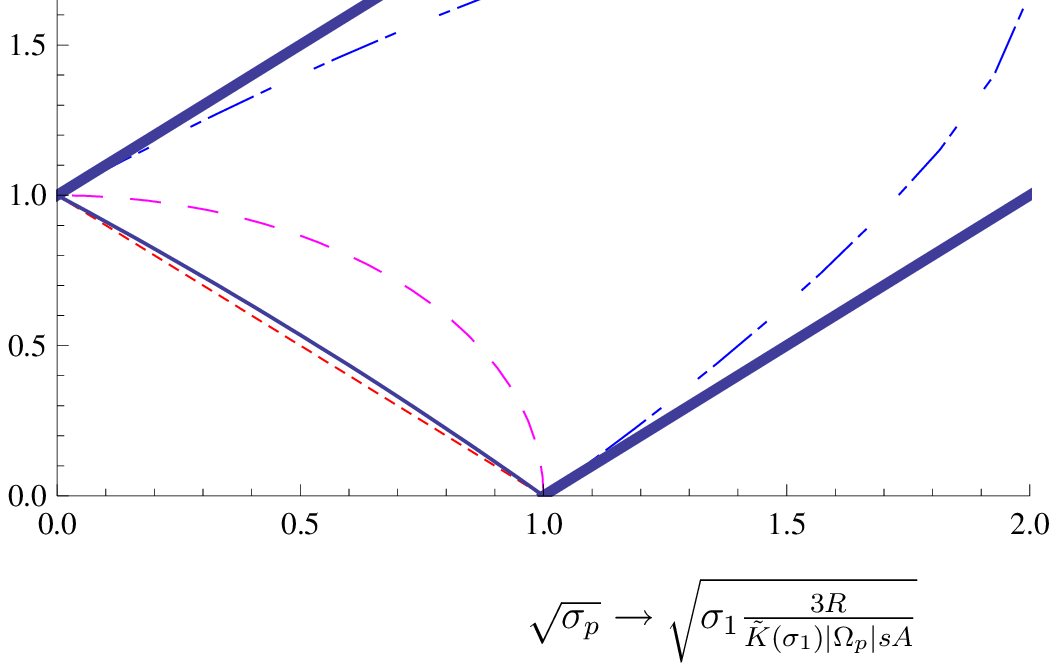}
}
 \subfloat[]
 {
\rotatebox{90}{\hspace{1.0cm}{$ \sqrt{\sigma_n} \rightarrow \sqrt{\sigma_1 \frac{3R}{\tilde{K}(\sigma_1)|\Omega_n |sA}}$}}
\includegraphics[height=0.20\textheight]{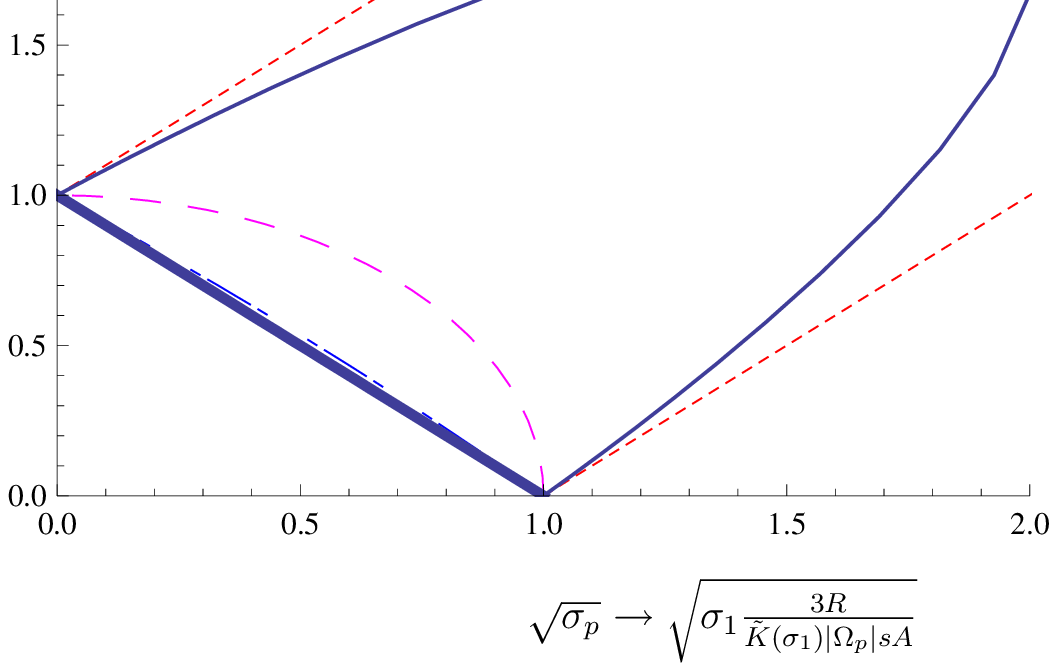}
}
\caption{ A "universal" exclusion plot in the $(\sqrt{\sigma_p},\sqrt{\sigma_n})$ plane exhibiting the dependence on the phase $\delta$. On te right panel the nuclear spins are of the same sign, while on the other of opposite signs.
 When the two amplitudes are relatively real, they are not bounded except when the the phase $\delta$ is the same with the relative phase of the two nuclear matrix elements.
In both panels  the
dotted,  the fine solid, the  dashed,  the dotted- dashed, and the thick solid curve
correspond to
 $\delta=0,\pi/6,\pi/2,5~\pi/6$ and $\pi$ respectively.
 \label{Fig:allparam}}
\end{center}
\end{figure}

\begin{figure}
\begin{center}
 \subfloat[]
 {
\rotatebox{90}{\hspace{1.0cm}{$ \sqrt{\sigma_n} \rightarrow \sqrt{\sigma_1 \frac{3R}{\tilde{K}(\sigma_1) s3}}$}}
\includegraphics[height=.20\textheight]{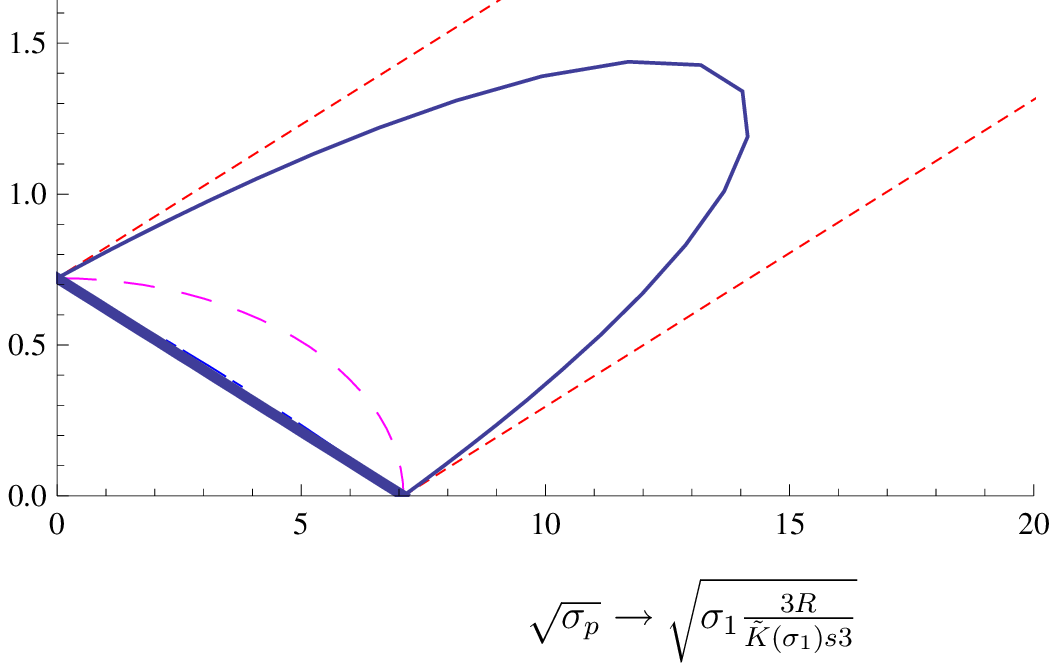}
}
 \subfloat[]
 {
\rotatebox{90}{\hspace{1.0cm}{$ \sqrt{\sigma_n}\rightarrow \sqrt{1.3 \times 10^{-4} \mbox{pb}}$}}
\includegraphics[height=0.20\textheight]{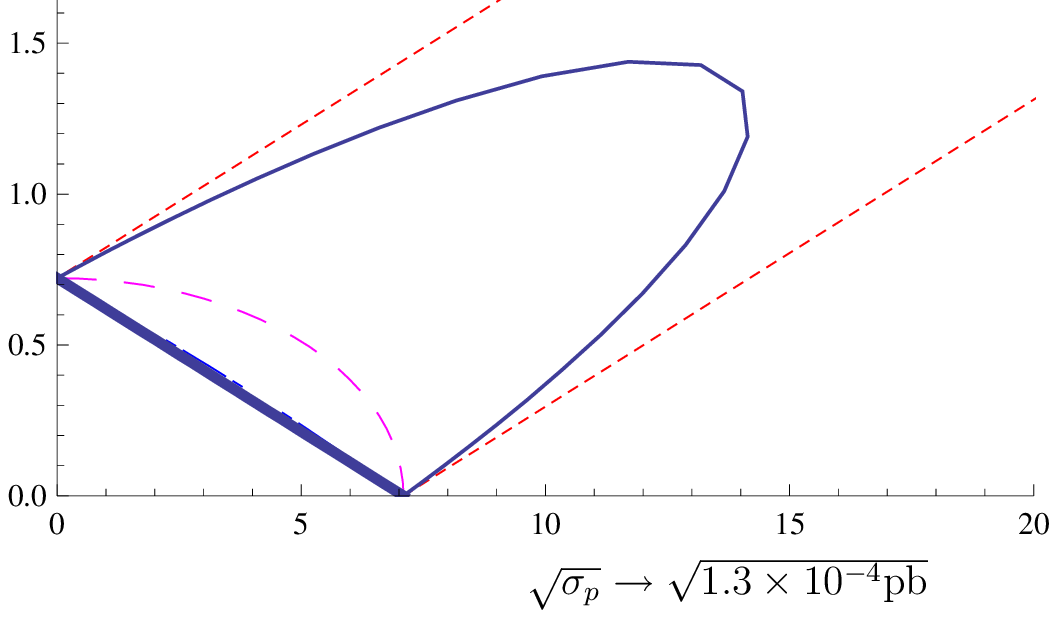}
}
\caption{ The exclusion plot in the $(\sqrt{\sigma_p},\sqrt{\sigma_n})$ plane in the case of the target $^3$He
for various values of the phase $\delta$ using the relevant spin ME of table \ref{table.spin}. The  units depend on  the parameters of table \ref{table.murt},
the experimental rate $R$ as well as  and, for the chosen scale of $\sigma_1$, on $\tilde{K}(\sigma_1)$ (a). 
The same exclusion plot in the case of a WIMP with a mass 20 GeV normalized to 1 event per kg target per year in the indicated units, obtained using $\tilde{K}(\sigma_1)=1.6\times 10^{2}$y$^{-1}$ and $\sigma_1=10^{-5}$pb  (b) . Cross sections for other event rates can be trivially extracted by a simple rescaling of panel (b).
 When the two amplitudes are relatively real ($\delta=0,\pi$), they are not bounded except when $\delta$ coincides with the phase difference $\delta_A$ of the neclear matrix elements. For the
 labelling of the curves see Fig: \ref{Fig:allparam}.
 \label{Fig:A3spin}}
\end{center}
\end{figure}

\begin{figure}
\begin{center}
 \subfloat[]
 {
\rotatebox{90}{\hspace{1.0cm}{$ \sqrt{\sigma_n} \rightarrow \sqrt{\sigma_1 \frac{3R}{\tilde{K}(\sigma_1) s19}}$}}
\includegraphics[height=.2\textheight]{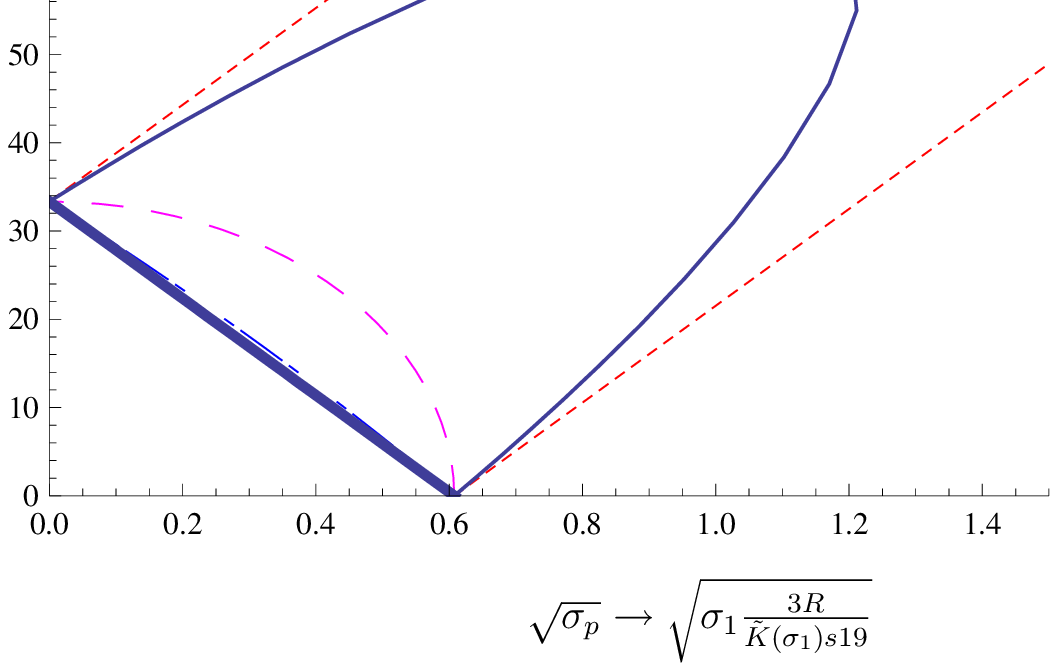}
}
 \subfloat[]
 {
\rotatebox{90}{\hspace{1.0cm}{$  \sqrt{\sigma_n}\rightarrow \sqrt{6.0 \times 10^{-5} \mbox{pb}}$}}
\includegraphics[height=0.2\textheight]{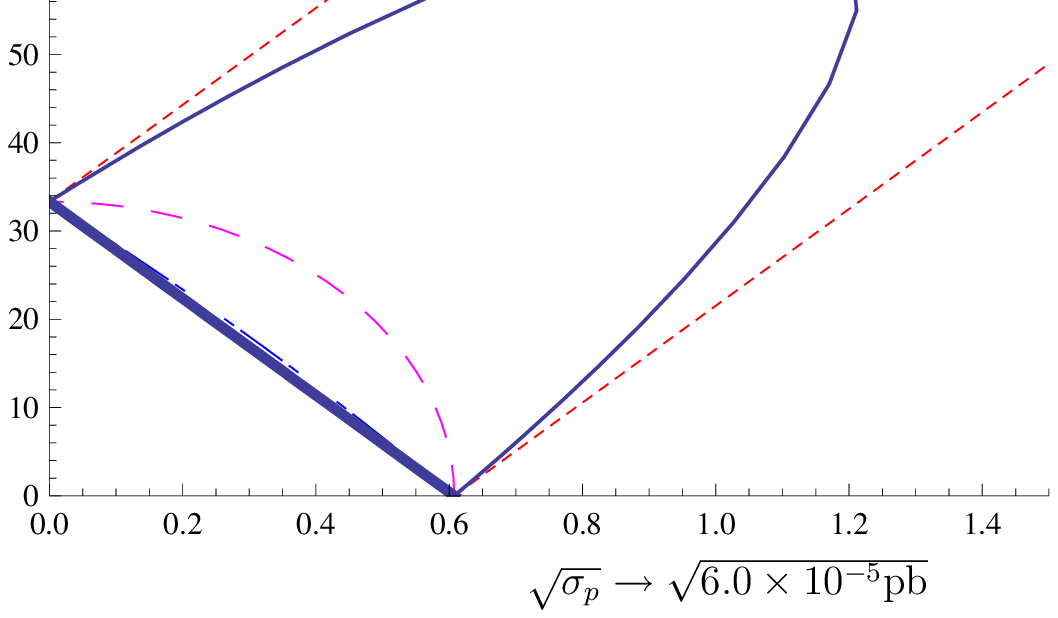}
}
\caption{ The same as in Fig. \ref{Fig:A3spin} in the case of $^{19}$F target.
 \label{Fig:A19spin}}
\end{center}
\end{figure}

\begin{figure}
\begin{center}
 \subfloat[]
 {
\rotatebox{90}{\hspace{1.0cm}{$  \sqrt{\sigma_n} \rightarrow \sqrt{\sigma_1 \frac{3R}{\tilde{K}(\sigma_1) s73}}$}}
\includegraphics[height=.20\textheight]{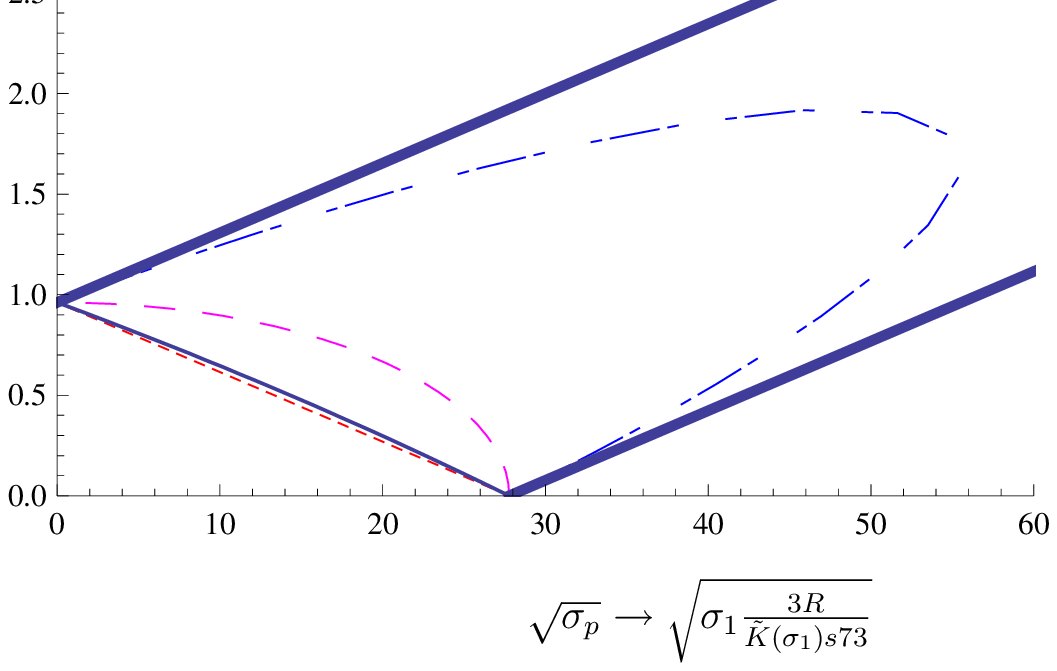}
\vspace{2.5cm}
}
 \subfloat[]
 {
\rotatebox{90}{\hspace{1.0cm}{$   \sqrt{\sigma_n}\rightarrow \sqrt{7.4 \times 10^{-5} \mbox{pb}}$}}
\includegraphics[height=0.20\textheight]{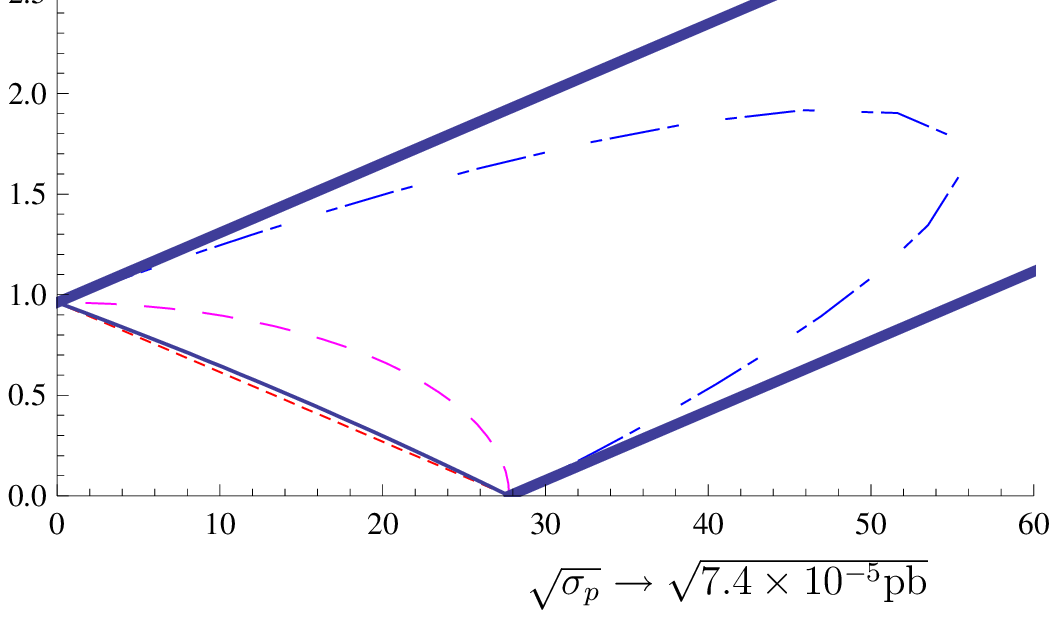}
}
\caption{ The same as in Fig. \ref{Fig:A3spin} in the case of $^{73}$Ge target. Now in panel (b) we exhibit
 the most sensitive case of a WIMP mass of 50 GeV.
 \label{Fig:A73spin}}
\end{center}
\end{figure}

\begin{figure}
\begin{center}
 \subfloat[]
 {
\rotatebox{90}{\hspace{1.0cm}{$  \sqrt{\sigma_n} \rightarrow \sqrt{\sigma_1 \frac{3R}{\tilde{K}(\sigma_1) s127}}$}}
\includegraphics[height=.20\textheight]{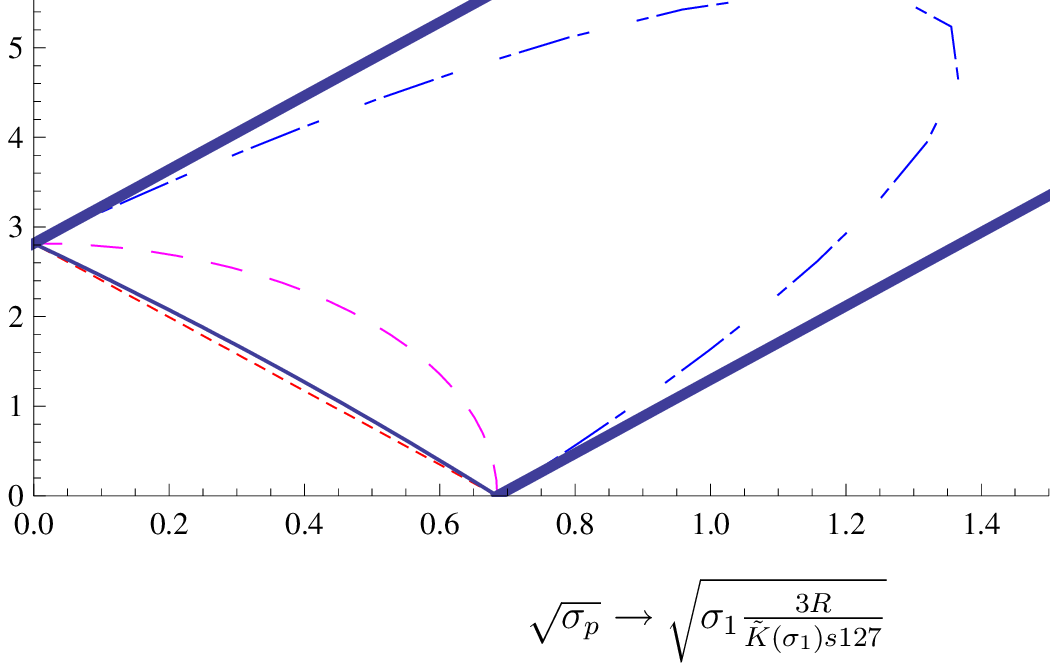}
}
 \subfloat[]
 {
\rotatebox{90}{\hspace{1.0cm}{$  \sqrt{\sigma_n}\rightarrow \sqrt{1.4 \times 10^{-4} \mbox{pb}}$}}
\includegraphics[height=0.20\textheight]{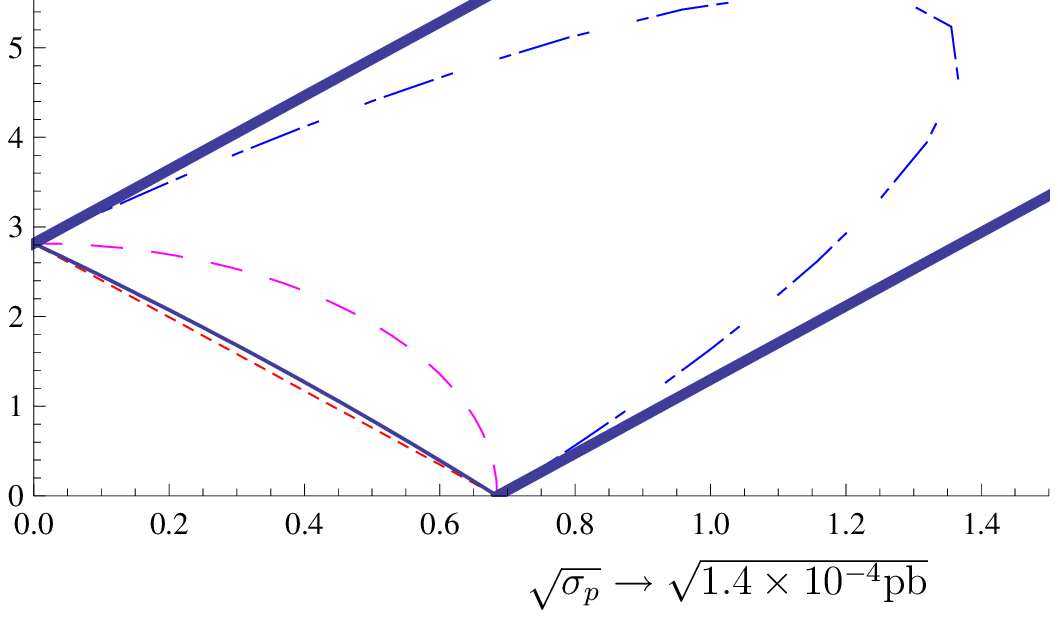}
}
\caption{ The same as in Fig. \ref{Fig:A3spin} in the case of $^{127}$I target. Now in panel (b) we exhibit
the most sensitive case of a WIMP mass of 30 GeV.
 \label{Fig:A127spin}}
\end{center}
\end{figure}

\begin{figure}
\begin{center}
 \subfloat[]
 {
\rotatebox{90}{\hspace{1.0cm}{$  \sqrt{\sigma_n} \rightarrow \sqrt{\sigma_1 \frac{3R}{\tilde{K}(\sigma_1) s23}}$}}
\includegraphics[height=.20\textheight]{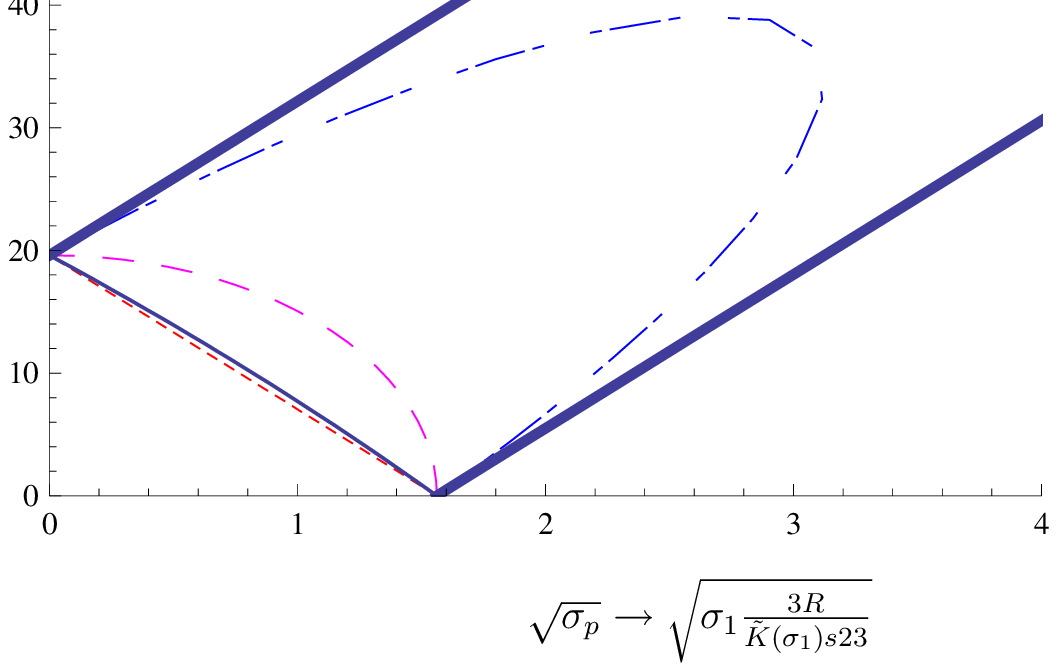}
}
 \subfloat[]
 {
\rotatebox{90}{\hspace{1.0cm}{$  \sqrt{\sigma_n}\rightarrow \sqrt{6.4 \times 10^{-5} \mbox{pb}}$}}
\includegraphics[height=0.20\textheight]{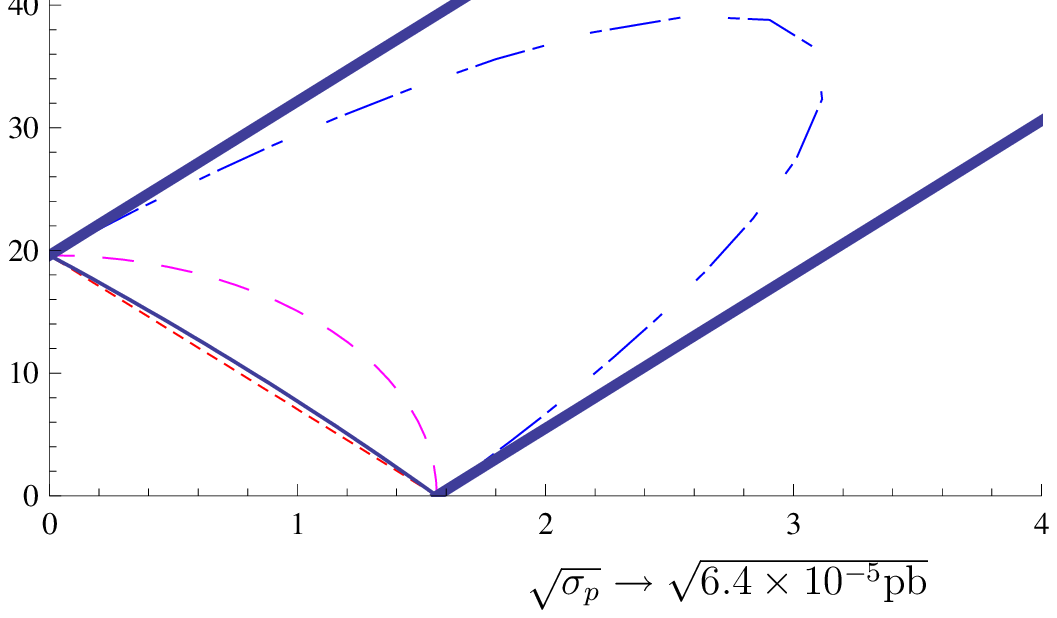}
}
\caption{ The same as in Fig. \ref{Fig:A3spin} in the case of $^{23}$Na target. Now in panel (b) we exhibit the most
sensitive case of a WIMP mass of 20 GeV.
 \label{Fig:A23spin}}
\end{center}
\end{figure}

 From the experimental data, using the nuclear spin matrix elements,one can extract a
restricted region in the  $\sigma_p,\sigma_n$ plane\cite{SGF05}-\cite{GIUGIR05}.
The relevant relation is:
\beq
||\Omega_p| \sqrt{ \sigma_p}+|{\Omega_n} |\sqrt{\sigma_n}
 e^{i \left(\delta+\delta_A\right )}|^2=\sigma_1 \frac{3R}{\tilde{K}(\sigma_1) sA}.
\eeq
where $sA$ is a short hand notation for $c_{spin}(A,\mu_r(A),m_{\chi^0})$.
The extracted values, given the event rate and the spin ME, depend on the WIMP mass and the relative phase of the two amplitudes.\\
Since the procedure is much more complicated than that entering
the analysis of the coherent node, a few explanations regarding the
presentation of our results (Figs
\ref{Fig:allparam}-\ref{Fig:A23spin}) are in order:
\begin{itemize}
\item We found it more convenient to present in the plots the extracted $\sqrt{{\sigma_p}}$ and $\sqrt{{\sigma_n}}$ rather than the cross sections themselves.
\item For illustrative purposes the dependence on $\delta$ can be given in a simple graph whereby the cross sections can be expressed in units
containing all the parameters. The extracted shapes, which  depend on $\delta_A$ are shown in
 Fig. \ref{Fig:allparam}. Such a plot, in principle, contains all the needed information, but it is too general to be  practical.
\item The contour for $\delta\ne0,\pi$ is in general an ellipse. For a given experimental bound, the allowed values of the cross sections
are in the space enclosed by an ellipse. One can see that, depending on $\delta$ the maximum allowed cross sections can be quite a bit higher than those extracted assuming  a single mode. If the two amplitudes are
relatively real, then the  contours become straight lines and
the cross sections may be constrained, but only if the relative phase of the two
amplitudes is the same with  $\delta_A$. If they differ by $\pi$, the individual cross
sections are not bounded, they can be anywhere between the two lines.
\item For given nuclear spin ME the extracted  $\sqrt{{\sigma_p}}$ and $\sqrt{{\sigma_n}}$ are presented in units
$\sqrt{\sigma_1 \frac{3R}{\tilde{K}(\sigma_1) sA)}}$ (see Fig.
\ref{Fig:A3spin} -\ref{Fig:A23spin} ). Once the experiment
determines the rate $R$ and the parameter $sA$, for the chosen
WIMP mass, is read off from table \ref{table.murt}, one can
immediately extract from the figures the cross sections in units
of $\sigma_1$ ( $\tilde{K}(\sigma_1)$ is given by Eq.
(\ref{Kconst})). As an illustration we do this on the right panel
of the the  figures \ref{Fig:A3spin} -\ref{Fig:A23spin} assuming
an event rate of 1 event per Kg target per year for the optimum
value of $m_{\chi}$ (maximum of $sA$)
\end{itemize}
 The following cases are of experimental interest:\\
i)  We first consider the case of  nuclear spin matrix elements of opposite sign and $|\Omega_n|>| \Omega_p|$
as is the case of the A=3 system. The exclusion plots are shown in Fig. \ref{Fig:A3spin}
\\ii) Next comes the case of spin matrix elements of opposite sign and $|\Omega_p|\gg |\Omega_n|$
as  is the case of the $^{19}$F target. This case has already been analyzed above, considering only protons. Just in case the elementary
 proton cross section is very suppressed, we present the relevant exclusion plots in Fig. \ref{Fig:A19spin}.
\\iii) After this we consider nuclear spin matrix elements of same sign and $|\Omega_n|\gg |\Omega_p|$.
This is the case of the $^{73}$Ge target. The relevant exclusion plots are shown  in Fig. \ref{Fig:A73spin}.
 \\iv We consider the  case with both spin matrix elements being significant.
Such  may be the case of the $^{127}$I target ($\Omega_p=1.127,\Omega_n=0.315$). The resulting exclusion
plots are shown in Fig. \ref{Fig:A127spin}.
\\v) We finally show the exclusion plots exhibited by another proton
favoring nucleus, $^{23}$Na, which is present together with
$^{127}$I in the target NaI (see Fig. \ref{Fig:A23spin}). We notice
that, since one has the same number of nuclei of each component in
a given mass of the target, $^{23}$Na  competes well with the
$^{127}$I in the spin induced event rate.

\section{Concluding Remarks}
We have analyzed the spin induced WIMP nucleus elastic cross section and related event rates. Both depend rather sensitively on the spin structure of the nucleus. Barring unusual circumstances at the elementary  level, the spin mode has no chance to compete with the coherent WIMP nucleus scattering in the case of heavy targets. It could, however, compete with it in the case of light targets. For light targets, and in particular for $^{3}$He and $^{19}$F, we believe the nuclear matrix elements are very accurate to allow reliable extraction of the nucleon cross sections from the data, if and when they become available. In the cases considered here, with the possible exception of $^{127}$I, the nuclear structure tends to favour the proton or the neutron component. This  allows a simple extraction of the corresponding nucleon cross section. This is also true even if  both components are present, but the isoscalar amplitude at the nucleon level is suppressed.  Finally, even if both the proton and the neutron amplitudes are important, we have shown that knowledge of the nuclear matrix elements allows one to draw suitable exclusion plots. Unfortunately, then, the situation is technically a bit complicated by the fact that one must draw one  exclusion plot for each WIMP mass. So, for targets with spin different from zero, exclusion plots should be drawn as more experimental data become available.
\section{Acknowledgments}
The final stages of this work were completed while the author visited Tuebingen under a  Humboldt Research Award.
The Author is indebted for this opportunity to Alexander von Humboldt Foundation and Professor Amand Faessler.

\end{document}